\documentclass{JHEP}
\usepackage{amsmath}
\usepackage{mathrsfs}

\title{Moduli Stabilization in Type IIB Flux Compactifications}
\author{Huan-Xiong Yang \\ Zhejiang Institute of Modern Physics, Physics Department,\\ Zhejiang
University, Hangzhou, 310027,
P. R. China \\E-mail: \email{hxyang@zimp.zju.edu.cn}}

\date{\today}

\abstract{In the present paper, we reexamine the moduli stabilization problem of the Type IIB orientifolds with
one complex structure modulus in a modified two-step procedure. The full superpotential including both the
3-form fluxes and the non-perturbative corrections is used to yield a F-term potential. This potential is
simplified by using one optimization condition to integrate the dilaton field out. It is shown that having
a locally stable supersymmetric Anti-deSitter vacuum is not inevitable for these orientifolds, which depend
strongly upon the details of the flux parameters. For those orientifolds that have stable/metastable
supersymmetry-broken minima of the F-term potential, the deSitter vacua might emerge even without the inclusion
of the uplifting contributions.
}


\begin{document}

One of the central topics in superstring phenomenology is studying the stabilization mechanism of compactification
moduli. In the context of Calabi-Yau compactification, the moduli fields generically include the dilaton,
the K\"{a}hler moduli and the complex structure moduli\cite{hep-th/9702155}. With the advent of DRS-GKP flux
stabilization mechanism\cite{hep-th/9908088,hep-th/0105097} and the CK-KKLT proposal for incorporating the possible
non-perturbative effects into the moduli stabilization scheme\cite{hep-th/0108220, hep-th/0301240}, much progress
has be made in this aspect, especially in understanding the moduli stabilization of Type IIB orientifold
compactification\cite{hep-th/0201028, hep-th/0308055, hep-th/0309187, hep-th/0402088, hep-th/0404257, hep-th/0407130,
hep-th/0411066, hep-th/0503072, hep-th/0503124, hep-th/0506090, hep-th/0506266}. The KKLT procedure\cite{hep-th/0301240}
has played a crucial guidance role in most of these achievements. However, argumentation\cite{hep-th/0411066,
hep-th/0506090, hep-th/0506266} does also exist on the validity of KKLT procedure that might be related to the question
if it is possible to have a stable vacuum with zero or positive cosmological constant in Type IIB string theory.

It was observed\cite{hep-th/9908088, hep-th/0105097} that, in the framework of Type IIB orientifold compactification,
turning on fluxes of NS-NS and RR 3-form gauge fields generates a no-scale type F-term potential for the complex
structure moduli ($U^i$) and the dilaton-axion field ($S$). Based on this observation, Kachru et al (KKLT) suggested
that to further freeze the K\"{a}hler moduli ($T^i$) the non-perturbative effects induced by Euclidean
D3-instantons\cite{hep-th/9604030} and/or by gaugino condensation in some hidden gauge group
sectors\cite{hep-th/0407130, hep-th/0503072} have to be taken into account. Similar proposal was also put forward by
Curio and Krause in studying the moduli stabilization of heterotic M-theory\cite{hep-th/0108220}. The KKLT proposal was
originally carried out in a two-step decoupled procedure (KKLT procedure) in which the dilaton-axion and the complex
structure moduli (if present) are assumed to be so heavy that they can be stabilized with only the 3-form fluxes. These
moduli are classically integrated out at first to create a constant superpotential $W_0$. Non-perturbative corrections
$W_{np}\sim g_ie^{-h_i T_i}$ from either D3-instantons or gaugino condensation effects to the superpotential are then
introduced to further stabilize K\"{a}hler moduli. The KKLT procedure appears to leads only to the string theory vacua
that are supersymmetric Anti-deSitter spaces. To reach an acceptable potential consistent with the present cosmological
observations, some uplifting mechanisms have to be made\cite{hep-th/0301240, hep-th/0309187} that promise to break
supersymmetry in some metastable vacua and allow a fine-tuning of the cosmological constant to a desired value.

Certainly being a promising scheme, there are still many further studies on the details of the context of the KKLT
procedure. It was shown in Refs.\cite{hep-th/0411066, hep-th/0506090, hep-th/0506266} that KKLT procedure does not
work for Type IIB orientifolds without complex structure moduli. By analyzing the stability properties of the associated
F-term potential, these authors concluded that the stable Anti-deSitter ground states are not possible if the orientifold
does not contain complex structure moduli (regardless of the number of untwisted K\"{a}hler moduli present in
$W_{np}$\cite{hep-th/0404257, hep-th/0403119}). Besides, de Alwis observed \cite{hep-th/0506266} that if the
non-perturbative corrections $W_{np}$ are included from the procedure beginning, there are extra terms in the resultant
F-term potential which are necessarily controlled by the same coefficients as the terms which are taken into account in
KKLT decoupled procedure. It cannot be justified in an acceptable approximation why these terms disappear in the original
KKLT scheme. The validity of D-term uplifting suggestion due to Kallosh et al\cite{hep-th/0309187} has also been commented
to be useless in the framework of KKLT two-step stabilization procedure, based on the relation
$2\Re f^{ab}D_b=\frac{ik^{ai}D_i W}{W}$ between the D-term and F-term potentials at a generic point in moduli space where
the superpotential is non-zero\cite{hep-th/0108200} (Here $k^{ai}$ stands for the generators of a Killing symmetry of the
K\"{a}hler metric and $f$ the gauge coupling function). Hence a supersymmetric Anti-deSitter ground state where all F-terms
vanish with $W \ne 0$ as in KKLT could not be lifted to a deSitter vacuum by adding a D-term potential. The alternative
uplifting suggestion proposed by KKLT themselves in Ref.\cite{hep-th/0301240}, in which the uplifting energy was attributed
to the interactions between D-brane and anti-D-branes, would be involved in an explicit supersymmetry breaking correction.
One has to search for viable uplifting mechanism in string cosmology studies because the explicit supersymmetry breaking
is generically out of control. In fact, it is possible to get metastable deSitter vacua without adding uplifting energies
within the KKLT proposal of the string moduli stabilization, if the full superpotential including both the flux contribution
and the non-superpotential corrections is considered throughout the stabilization procedure\cite{hep-th/0506090}. An
illustrative example of the metastable deSitter minima for models with one complex structure modulus has been given by de
Alwis in the light $T$ approximation\cite{hep-th/0506266}.

In this paper, we reexamine the moduli stabilization problem of Type IIB orientifolds with just one complex structure
modulus in their orbifold limits. Instead of using the problematic KKLT procedure,  we adopt an alternative two-step
procedure that has its roots in the confirmed one-step procedure developed by L\"{u}st et al\cite{hep-th/0506090}. The
distinction between our method and that in Ref.\cite{hep-th/0506266} lies on the fact that we do not take the light
K\"{a}hler moduli approximation. What we have found is that the stable/metastable supersymmetric Anti-deSitter vacua are
only accessible for some of these orientifolds. The criteria are given for making judgement. Among those orientifolds
that have no stable/metastable supersymmetric ground states, there are some models whose F-term potentials have deSitter
minima, very attractive for applications in brane cosmology.

The models we consider here are Type IIB orientifolds with Hodge numbers $h^{untw}_{(1,1)}=3$ and $h^{untw}_{(2,1)}=1$ in
their untwisted moduli spaces\cite{hep-th/0506090}. These orientifolds are assumed to be compactified on the toroidal
orbifolds $X_6=T^6/{\Gamma}$ with the orbifold groups $\Gamma={\bf Z}_{6-II}$, ${\bf Z}_2 \times {\bf Z}_3$ and
${\bf Z}_2 \times {\bf Z}_6$\cite{hep-th/9804026, hep-th/9708040}. As usually done in literature, the untwisted K\"{a}hler
moduli and the untwisted complex structure moduli of the orientifolds are labeled by $T^i~(i=1,2,3)$ and $U$ respectively.
For simplicity we concentrate on the isotropic case in which $T^1=T^2=T^3=T$. The K\"{a}hler potential of these models
reads\cite{hep-th/0506090}
\begin{equation}
K=-3 \ln (T +\bar{T}) -\ln {S + \bar{S}} - \ln (U + \bar{U})
\label{Eq: 1}
\end{equation}
The superpotential consists of two terms $W=W_{flux}+ 3g e^{-hT}$ where $W_{flux}$ stands for the contribution of the
3-form fluxes and $3g e^{-hT}$ the possible non-perturbative correction. The prefactor $g$ of the non-perturbative term
is assumed to be a constant, reflecting the ignorance of the probable perturbative corrections in our discussion. The
relevant 3-forms on the orientifolds are
$\omega_{A_0} = dz^1 \wedge dz^2 \wedge dz^3$,
$\omega_{A_3} = dz^1 \wedge dz^2 \wedge d{\bar z}^3$,
$\omega_{B_0} = d\bar{z}^1 \wedge d\bar{z}^2 \wedge d\bar{z}^3$
and
$\omega_{B_3} = d\bar{z}^1 \wedge d\bar{z}^2 \wedge dz^3$. In terms of these complex 3-forms, the flux $G_3=F_3 +iS H_3$
is expanded as
\begin{equation}
\frac{G_3}{(2 \pi)^2 \alpha^{\prime}}=A^0(S, U)\omega_{A_0} + A^3(S, U)\omega_{A_3} +B^0(S, U)\omega_{B_0}
+B^3(S, U)\omega_{B_3}~
\label{Eq: 2}
\end{equation}
where the complex coefficients take the form
\begin{equation}
\left.
\begin{array}{ll}
A^0(S, U)=A^0_1(U) + i A^0_2(U)S,~ ~ ~ ~ A^3(S, U)=A^3_1(U) +i A^3_2(U)S ~\\
B^0(S, U)=B^0_1(U) -i B^0_2(U)S,~ ~ ~ ~ B^3(S, U)=B^3_1(U) -i B^3_2(U)S ~
\end{array}
\right.
\label{Eq: 3}
\end{equation}
and $A^{0,3}_{1,2}(U)$ and $B^{0,3}_{1,2}(U)$ each contain a constant term and a term linear in $U$. All together they
comprise eight real integer-valued flux parameters, whose explicit forms depend on the geometric details of each
individual orientifold\cite{hep-th/0506090}. The flux-related superpotential which is defined by
$\dfrac{1}{(2\pi)^2 \alpha^{\prime}} \int_{CY_3} G_3 \wedge \Omega$ \cite{hep-th/0105097}\cite{hep-th/0506090} is
expressed as
\begin{equation}
W_{flux}= B^0_1(U) -i B^0_2(U) S
\label{Eq: 4}
\end{equation}
Taking into account of the possible non-perturbative corrections, the full superpotential for the present models is
$W = \lambda \Big(B^0_1(U) -i B^0_2(U) S\Big)+3ge^{-hT}$~(Here the parameter $\lambda$ is used to reflect the relative
ratio between the flux contribution to the superpotential and the one from the non-perturbative corrections). Hence,
for Type IIB orientifolds with just one complex structure modulus, the full (untwisted) superpotential has a form as
follows
\begin{equation}
W = \alpha_0 + \alpha_1 U + \alpha_2 S + \alpha_3 S U + 3 g e^{-h T}
\label{Eq: 5}
\end{equation}
where $\alpha_i~(i=0,~1,~2,~3)$ are some real flux parameters.  To ensure the 3-form fluxes dominating over the
superpotential we assume $h>0$ in Eq.(\ref{Eq: 5}). We also assume $g>0$ for concreteness. Since the F-term
potential (See below) is invariant under the reversal transformation $W \rightarrow -W$ of the full superpotential,
taking $g>0$ does not bring out any loss of generality. If $\alpha_i~(i=1,~2,~3)$ vanish, Eq.(\ref{Eq: 5}) will be
reduced to the same expression as the superpotential employed in Ref.\cite{hep-th/0301240} to fix the K\"{a}hler
moduli. In the original KKLT procedure such a superpotential would emerge after the complex structure modulus (including
the dilaton-axion field) were fixed solely by the 3-form flux effects. In the one-step procedure\cite{hep-th/0506090},
however,  this special superpotential means that the complex structure modulus and dilaton-axion field are completely
free in the corresponding models. Moreover, the field $U$ (or $S$) will escape from being fixed if $\alpha_1$
(or $\alpha_2$) vanishes. Because we are exclusively interested in the models in which all compactification moduli
could be stabilized, we assume $\alpha_1 \ne 0$ and $\alpha_2 \ne 0$ in what follows.

In terms of the language of $\mathscr{N}=1$ supergravity\cite{Wess}, the potential energy of the considered models
can be organized into the standard form of the F-term potential\cite{hep-th/0105097}
\begin{equation}
V_F = e^K (|D_S W|^2 + |D_U W|^2 + 3|D_T W|^2 -3|W|^2 )
\label{Eq: 6}
\end{equation}
with $D_i W$ the K\"{a}hler derivatives of superpotential with respect to the moduli fields,
$D_i W= \partial_i W + W \partial_i K~(i=S, U, T)$. We express the moduli fields as
$T=t+i\tau$, $S=s+i\sigma$ and $U=u+i\nu$. The real moduli fields $t$, $u$ and $s$ should take positive values, as
implied by the K\"{a}hler potential (\ref{Eq: 1}). The points $\tau = \sigma = \nu =0$ define some flat directions
in moduli space on which  $\partial_{\tau}V_F=\partial_{\sigma}V_F=\partial_{\nu}V_F=0$. We confine ourselves to these
points, at which the F-term potential has a simple expression:
\begin{equation}
\left.
\begin{array}{lll}
V_F & =  \frac{1}{16 t^3 s u} & \Big[ 6g ht e^{-ht}  ( \alpha_0 + u \alpha_1 + s \alpha_2
+  u s \alpha_3 + 3g e^{-ht} ) + 6 ( g h t)^2 e^{-2ht} \\
&   & + ( \alpha_0 - u s \alpha_3 + 3 g e^{-h t} )^2 + (u \alpha_1 - s \alpha_2)^2 \Big]
\end{array}
\right.
\label{Eq: 7}
\end{equation}
Potential (\ref{Eq: 7}) is our main concern in this paper.

It follows from Eq.(\ref{Eq: 7}) that the remaining optimization conditions
$\partial_{t}V_F=\partial_{s}V_F=\partial_{u}V_F=0$ demand
\begin{eqnarray}
&  &g^2 \big[ 9+18 ht + 14 (h t)^2 + 4(h t)^3 \big] + 2g e^{ht} \big[(3 + 3 ht + (h t)^2)\alpha_0
+ ht (2u\alpha_1 + 2s\alpha_2 + su\alpha_3)   \nonumber \\
&  &  ~ ~ ~ ~ ~ ~ ~ ~  + (h t)^2 (u\alpha_1 + s\alpha_2 + su\alpha_3)
- 3 s u \alpha_3 \big] + e^{2ht} \big[(\alpha_0 -su\alpha_3)^2 + (u\alpha_1 - s\alpha_2)^2 \big] = 0 \nonumber \\
&  &3g^2 \big[ 3 + 6ht + 2(h t)^2 \big] + 6g e^{ht} \big[ \alpha_0 (1+ ht) + u \alpha_1 ht \big]+ e^{2ht} \big[ \alpha_0^2
+ (u \alpha_1)^2 -(s \alpha_2)^2 -(su\alpha_3)^2 \big] =0  \nonumber \\
&  &3g^2 \big[ 3 + 6ht + 2(h t)^2 \big] + 6g e^{ht} \big[ \alpha_0 (1+ ht) + s \alpha_2 ht \big]+ e^{2ht} \big[ \alpha_0^2
- (u \alpha_1)^2 +(s \alpha_2)^2 -(su\alpha_3)^2 \big] =0 \nonumber
\end{eqnarray}
\begin{equation}
\label{Eq: 8}
\end{equation}
In the standard one-step procedure, the moduli fields $t$, $u$ and $s$ should be "fixed" simultaneously by solving
the above equations. L\"{u}st et al have got a special solution to Eqs.(\ref{Eq: 8}) in this spirit, which
describes the moduli stabilization mechanism for models at whose F-term potential minima the $\mathscr{N}=1$
supersymmetry is finally restored\cite{hep-th/0506090}. The analysis directly on the basis of one-step procedure
is in general complicated. Here we adopt an alternative instead. The structural similarity of the last two equations
in Eqs.(\ref{Eq: 8}) implies that if the complex structure modulus $u$ is "freezed" the dilaton field $s$ will be
"freezed" through either $\alpha_2 s = \alpha_1 u$ or $\alpha_2 s = -\alpha_1 u -3ght e^{-ht}$. In what follows we assume
that the dilaton $s$ has been integrated out by applying either of these two constraints. We do not choose to integrate
out both $s$ and $u$ as did in original KKLT procedure because by solving only the last two equations in Eqs.(\ref{Eq: 8})
(i.e., $\partial_s V_F=\partial_u V_F=0$) it is impossible to express these two moduli in terms of the K\"{a}hler modulus
$t$. Firstly integrating the dilaton $s$ out in the present procedure does not mean the congealment of this modulus
at once. Similar to the one-step procedure, in our procedure all the moduli fields are expected to be fixed (if possible)
simultaneously. Because the moduli fields $t$, $u$ and $s$ must take positive values, $\alpha_2 s = \alpha_1 u$ implies
that in the corresponding models the flux parameters $\alpha_1$ and $\alpha_2$ have the same sign while
$\alpha_2 s = -\alpha_1 u -3ght e^{-ht}$ implies that $\alpha_1$ and $\alpha_2$ have the opposite sign.

We first consider the models in which $\alpha_1$ and $\alpha_2$ have the same sign. After the dilaton field $s$ is
integrated out by using $\alpha_2 s = \alpha_1 u$, the F-term potential of these models can be recast as
\begin{equation}\label{Eq: 9}
    V_F = \frac{\alpha_2}{16 \alpha_1 t^3 u^2} \bigg[ 6 (ght)^2 e^{-2ht} + 6 ght e^{-ht} (\alpha_0
    + 2\alpha_1 u + \frac{\alpha_1 \alpha_2 u^2}{\alpha_2} + 3 g e^{-ht})
    + (\alpha_0 - \frac{\alpha_1 \alpha_3 u^2}{\alpha_2} + 3 g e^{-ht})^2 \bigg]
\end{equation}
The conditions which have to be fulfilled at the critical points of potential (\ref{Eq: 9}) are either
\begin{equation}
\alpha_3 = \frac{\alpha_1 \alpha_2 (\alpha_0 + 3ge^{-ht} )}{\big[ \alpha_0 + (3 +ht)g e^{-ht} \big]^2}
~~~~~~~~
u =- \frac{\alpha_0 + (3+ ht)ge^{-ht}}{\alpha_1}
\label{Eq: 10}
\end{equation}
or
\begin{eqnarray*}
& &\alpha_3=-\frac{\alpha_1 \alpha_2 (g h t)^2 (7+2ht)^3  \big[ 3g(1+ 4ht + 2h^2t^2)e^{-ht}
+ (1+2ht)\alpha_0  \big]}{\big[ g^2 (72  + 177 h t +109h^2t^2 +16h^3t^3 -2h^4t^4 )e^{-ht}
+ g(48 + 71 h t +22 h^2t^2)\alpha_0
+4 (2+ h t) \alpha_0^2 e^{ht}  \big]^2 } \\
& & ~~~\\
& &u = -\frac{g^2(72 +177ht +109h^2t^2 +16h^3t^3 -2h^4t^4)e^{-h t} +g(48 +71h t +22h^2t^2)\alpha_0
+4(2+h t)\alpha_0^2 e^{ht}}{g h t(7+2ht)^2 \alpha_1}
\end{eqnarray*}
\begin{equation}\label{Eq: 11}
\end{equation}
At a critical point defined by Eqs.(\ref{Eq: 10}), the $\mathscr{N}=1$ supersymmetry is restored,
$D_T W = D_S W = D_U W =0$\cite{hep-th/0506090}, and the potential takes a definite-negative critical value,
\begin{equation}\label{Eq: 12}
V_F^{susy}=-\frac{3\alpha_1 \alpha_2 g^2 h^2 e^{-2ht}}{8t \Big[ \alpha_0 + (3+ht)ge^{-ht} \Big]^2}
\end{equation}
At a critical point defined by Eqs.(\ref{Eq: 11}), on the other hand, the supersymmetry is spontaneously broken. In
this case, the critical value of F-term potential (\ref{Eq: 9}) which is found to be $V_F^{non-susy}=A/B$  with
\begin{eqnarray}\label{Eq: 13}
    A& = & \alpha_1 \alpha_2 (gh)^2 (7 + 2ht)^2 \Big[ 3g^2 (-48 -84ht -73h^2t^2 -22h^3t^3 +2h^4t^4) \nonumber \\
     &   & + 12 g(-8 -5ht +2h^2t^2 + 2h^3t^3)\alpha_0 e^{ht} \nonumber \\
     &   & + 8 (-1+ht)(2+ht)\alpha_0^2 e^{2ht} \Big]
\end{eqnarray}
and
\begin{eqnarray}\label{Eq: 14a}
    B& = & 8t \Big[ 16(2+ht)^2 \alpha_0^4 e^{4ht} + 8g(2+ht)(48+71ht+22h^2t^2)\alpha_0^3 e^{3ht} \nonumber \\
     &   & +g^2 (-3456 -10224ht-10313h^2t^2 -4252h^3t^3 -580h^4t^4 +16h^5t^5)\alpha_0^2 e^{2ht} \nonumber \\
     &   & +2g^3 (48 + 71ht +22h^2t^2)(-72 -177ht-109h^2t^2 -16h^3t^3 +2h^4t^4)\alpha_0 e^{ht} \nonumber \\
         &   & +g^4(-72 -177ht-109h^2t^2 -16h^3t^3 +2h^4t^4)^2    \Big]
\end{eqnarray}
is not necessary to be negative-definite. In general, a Type IIB orientifold may have both supersymmetry-preserving
and supersymmetry-broken crises at the same time. Provided $\alpha_0 \geq 0$, nevertheless, the supersymmetric
crises are only possible for models with $\alpha_1 <0$, $\alpha_2<0$ and $\alpha_3>0$ while the supersymmetry-broken
crises defined by Eqs.(\ref{Eq: 11}) are only accessible for models with $\alpha_3 <0$.

To determine whether the obtained crises are local minima of the F-term potential (\ref{Eq: 9}), we have to calculate
the second order derivatives of potential (\ref{Eq: 9}) with respect to the moduli fields $t$ and $u$ and then
define the so-called Hessian determinants. Let $D_{tt}={\partial^2 V_F}/{\partial t^2}$,
$D_{uu}={\partial^2 V_F}/{\partial u^2}$, $D_{ts}={\partial^2 V_F}/{{\partial t}{\partial u}}$ and
$\Delta =D_{tt}D_{uu}-D^2_{tu}$. At the supersymmetry-restoring crises defined by Eqs.(\ref{Eq: 10}),  these Hessian
determinants are found to be
\begin{eqnarray}
&   & \Delta = \frac{3g^2h^2 \big[(1+2ht)\alpha_0 + 3(1+ht)ge^{-ht} \big]\big[ 2(2+ht)\alpha_0
+ 3(4+3ht+h^2t^2)ge^{-ht}\big]\alpha_1^4\alpha_2^2 e^{-6ht}}{32t^6 \big[ \alpha_0 + (3+ht)ge^{-ht}\big]^6}  \nonumber \\
&   & D_{tt}=\frac{3g^2h^2(5+5ht+2h^2t^2)\alpha_1 \alpha_2 e^{-2ht}}{8t^3\big[ \alpha_0 + (3+ht)ge^{-ht} \big]^2}
\label{Eq: 15a}
\end{eqnarray}
Being a local minimum for the critical potential (\ref{Eq: 12}) requires both $D_{tt}$ and $\Delta$ being positive.
Therefore, that a supersymmetric crisis becomes a local minimum is possible only if the $t$-coordinate of the
critical point falls into the intervals $\alpha_0 > - \frac{3(1+ht)ge^{-ht}}{1+2ht}$ or
$\alpha_0 < - \frac{3(4+3ht+h^2t^2)ge^{-ht}}{2(2+ht)}$. The sample model
\begin{math}
    W=b_2 -d_2 S-\frac{a_0}{2}U+\frac{c_2}{\sqrt 3} SU + ge^{-ht}
\end{math}
provided by L\"{u}st et al in Eq.(3.50) of Ref.\cite{hep-th/0506090} with all coefficients $a_0$, $b_2$, $c_2$ and $d_2$
positive is very a special case where the first inequality holds. If the $t$-coordinate of the critical point lies in
the interval $- \frac{3(4+3ht+h^2t^2)ge^{-ht}}{2(2+ht)} < \alpha_0 < - \frac{3(1+ht)ge^{-ht}}{1+2ht}$, the supersymmetric
crisis is only a saddle point. As an illustration to this exceptional situation we consider a toy model with superpotential
\begin{math}
    W \approx -1.2\times 10^{-5} -2.5\times 10^{-3}S -2.5\times 10^{-3}U -3.37535SU + 3e^{-T}
\end{math}. The potential curve has a supersymmetric crisis with coordinates $t\approx 13.860922$ and $u\approx 0.001645$
in moduli space. This $t$-coordinate does just fall into the interval in which the latter inequalities holds. The second
Hessian determinant is found to take a negative value $\Delta \approx -5.54\times 10^{-11}$ at this supersymmetric crisis,
as a result, this critical point is neither a local minimum nor a local maximum. In fact, the F-term potential of this
toy model has a supersymmetry-broken minimum $V_F^{Min} \approx -3.47\times 10^{-8}$ (in $M_p =1$ units) at the point with
coordinates $t \approx 10.046915$ and $u \approx 0.018511$, at which the corresponding Hessian determinants take positive
values $D_{tt} \approx 4.72\times 10^{-7}$ and $\Delta \approx 1.88 \times 10^{-10}$. Although such a local minimum is an
Anti-deSitter space, it can be uplifted to the desired deSitter space in a controllable manner by including the possible
D-term contributions in the full potential energy\cite{hep-th/0108200}.

Is it possible to obtain a stable/metastable deSitter vacuum for some Type IIB orientifold with one complex structure modulus
( in its untwisted sector ) directly from its F-term potential?  The answer is absolutely yes. To have a simple
confirmation to this answer, we focus on the models with $\alpha_0=0$. For such a model to have a deSitter vacuum that is
bound to be a supersymmetry-broken crisis, the flux parameter $\alpha_3 $ in its superpotential must be negative. The
position (in moduli space) of this possible deSitter vacuum is determined by Eqs.(\ref{Eq: 11}), at which the potential
energy and the Hessian determinants turn out to be
\begin{equation}
    V_F^{\textrm{NonSusy}}=\frac{3\alpha_1\alpha_2 h^2(7+2ht)^2(-48-84ht-73h^2t^2-22h^3t^3+2h^4t^4)}
    {8t(-72-177ht-109h^2t^2-16h^3t^3+2h^4t^4)^2}
    \label{Eq: 16}
\end{equation}
and
\begin{eqnarray}
& D_{tt}=     & \frac{3\alpha_1 \alpha_2 h^2 (7+2ht)^2 (288+792ht+869h^2t^2+547h^3t^3+234h^4t^4+58h^5t^5+8h^6t^6)}
{8t^3(-72-177ht-109h^2t^2-16h^3t^3+2h^4t^4)^2} \nonumber \\
& \Delta =    & \frac{9\alpha_1^4 \alpha_2^2 h^6 (7+2ht)^8}{32g^2t^2(-72-177ht-109h^2t^2-16h^3t^3+2h^4t^4)^6} \nonumber \\
             &   & ~ ~ ~ ~ ~ \cdot \bigg[-2(72+105ht+128h^2t^2+117h^3t^3+54h^4t^4+10h^5t^5)^2  \nonumber \\
             &   & ~ ~ ~ ~ ~ ~ + (2+ht)(39+93ht+68h^2t^2+22h^3t^3) \nonumber \\
         &   & ~ ~ ~ ~ ~ ~ \cdot (288+792ht+869h^2t^2+547h^3t^3+234h^4t^4+58h^5t^5+8h^6t^6)\bigg]
\label{Eq: 17}
\end{eqnarray}
respectively. Since $\alpha_1$ and $\alpha_2$ take the same sign, $D_{tt}$ is always positive. If $\Delta$ takes a positive
value further, the corresponding potential extremum in Eq.(\ref{Eq: 16}) will be a local minimum. Such a minimum is
manifestly unnecessary to be negative. A deSitter vacuum is accessible for such a model if the $t$-coordinate of its
critical point obeys the following inequalities:
\begin{eqnarray}
&   & -2(72+105ht+128h^2t^2+117h^3t^3+54h^4t^4+10h^5t^5)^2 \nonumber \\
&   & ~ ~ ~ ~ ~ ~ + (2+ht)(39+93ht+68h^2t^2+22h^3t^3) \nonumber \\
&   & ~ ~ ~ ~ ~ ~ \cdot (288+792ht+869h^2t^2+547h^3t^3+234h^4t^4+58h^5t^5+8h^6t^6)>0~, \nonumber \\
&   & -48-84ht-73h^2t^2-22h^3t^3+2h^4t^4>0~.
    \label{Eq: 18}
\end{eqnarray}
In other words, when such a model reaches its deSitter vacuum, the K\"{a}hler modulus $t$ will be freezed at the interval
specified by inequalities (\ref{Eq: 18}). This interval can be numerically approximated as
$\frac{13.860922}{h} <t <\frac{14.697531}{h}$. A sample model with superpotential
$$
    W=S+U-1.67539\times 10^{8} SU + 3e^{-T}
$$
is given for illustration. The model has a local deSitter minimum $V_F^{\textrm{Min}}\approx 3.76\times10^{-4}$ (in the
units of $M_p=1$) at a point in moduli space with coordinates $t=14$ and $u\approx 4.3705\times 10^{-7}$ (So $s \approx
4.3705\times 10^{-7}$). As required, both the Hessian determinants at this minimum are positive-definite,
$D_{tt}\approx 0.2104$ and $\Delta \approx 7.76\times 10^9$.

A brief remark follows now on the moduli stabilization of the Type IIB orientifolds in which the flux parameters $\alpha_1$
and $\alpha_2$ have opposite sign. For these models, the F-term potential becomes
\begin{eqnarray}
    V_F = &  - \frac{1}{16t^3u\alpha_2(u\alpha_1 + 3ghte^{-ht})} \bigg[ & 3\alpha_2^2 g^2e^{-2ht}(3+6ht-h^2t^2)
    +6\alpha_2^2ge^{-ht}(\alpha_0 +ht\alpha_0 +2htu\alpha_1) \nonumber \\
    &  & ~ + \alpha_2^2(\alpha_0^2 + 4u^2 \alpha_1^2)+ u^2\alpha_3^2 (u\alpha_1 + 3ghte^{-ht})^2  \nonumber \\
    &  & ~ + 2u\alpha_2\alpha_3 (\alpha_0 + 3g^{-ht} -3ghte^{-ht})(u\alpha_1 +3ghte^{-ht}) \bigg]
    \label{Eq: 19}
\end{eqnarray}
after the dilaton $s$ is integrated out ( by employing the constraint $\alpha_2 s = -\alpha_1 u -3ght e^{-ht}$ )
in our two-step procedure. If $\alpha_1<0$, this potential has a crisis $(t,~u)$ satisfying the following equations
\begin{eqnarray}
    &   &u=-\frac{3ghte^{-ht}}{2\alpha_1}~, \nonumber \\
    &   &16\alpha_1^2\alpha_2^2 \Big [ 9g^2(-5+2ht)(1+2ht)e^{-2ht} +6\alpha_0 ge^{-ht}(-5-3ht+h^2t^2) + \alpha_0^2
    (-5+2ht) \Big ] \nonumber \\
    &   &~ ~ + 216\alpha_1\alpha_2\alpha_3 g^2h^2t^2e^{-2ht} \Big[\alpha_0 + (3-ht-h^2t^2)ge^{-ht} \Big] \nonumber \\
    &   &~ ~ - 81\alpha_3^2(1+2ht)g^4h^4t^4e^{-4ht} =0~.
    \label{Eq: 20}
\end{eqnarray}
However, no matter what the derivative $D_{tt}$ is, the second Hessian determinant $\Delta$ does always vanish at the
crisis. This disables us from making a simple and definite judgement on whether the corresponding crisis is an extremum
of the F-term potential, although such a possibility is not excluded.

In conclusion, we have reexamined the moduli stabilization problem in the Type IIB orientifolds with one complex
structure modulus. Our investigation is essentially based on the CK-KKLT mechanism that the K\"{a}hler moduli
$T^i$ can be stabilized at the string vacuum by non-perturbative effects. We start from the full superpotential
that includes the both contributions of the $T^i$-independent 3-form fluxes and the $T^i$-dependent non-perturbative
corrections. Nevertheless, our procedure is still a two-step one. Although we do not use the light $T^i$
approximation\cite{hep-th/0506266} to stabilize the heavy moduli in the first stage, we use one optimization condition
to integrate out the dilaton field firstly. This two-step procedure has its roots in the more stringent one-step
procedure\cite{hep-th/0506090} but much simpler in practice. What we have found is that the metastable supersymmetric
Anti-deSitter vacua are unnecessarily accessible for some Type IIB orientifolds with one complex structure modulus.
Whether a model has a supersymmetry preserving Anti-deSitter vacuum depends greatly upon the choice of flux parameters
in the superpotential. In view of the potential applications in phenomenology, the orientifolds that have no supersymmetric
Anti-deSitter vacua appear more attractive. Some of these models that possess the deSitter-like vacua (with positive
energy minima) even at the level of F-term potential are expected to form a reliable platform for studying K\"{a}hler
moduli inflation\cite{hep-th/0509012}. Because the supersymmetry-broken F-term potential minima can be uplifted to the
deSitter vacua through introduction of the movable D7-branes into the orientifold configuration, these
models do also provide a viable scenario in a more extensive sense for realizing D-term inflation in string theory. We
stress here again that the above results are obtained within the supergravity approximation that might strongly depend
upon the details of the K\"{a}hler potential and superpotential of the Type IIB orientifolds with just one complex
structure modulus.

\section*{Acknowledgments}
The author is grateful to L. Zhao and W. Schulgin for useful discussions. This work is partially supported by CNSF-10375052,
CNSF-10247009 and the Startup Foundation of the Zhejiang Education Bureau.

\bibliographystyle{JHEP}
\bibliography{Moduli}
\end{document}